\documentclass[aps,pra,reprint,superscriptaddress,footinbib,longbibliography]{revtex4-2}

\usepackage{amssymb}
\usepackage{amsmath}
\usepackage{amsfonts}
\usepackage{xcolor}
\usepackage{graphicx}
\usepackage{bm}
\usepackage{physics}
\usepackage{mathrsfs} 
\usepackage[unicode]{hyperref}
\usepackage{comment}

\hypersetup{
   unicode=true,          
   plainpages=false,
   colorlinks=true,       
   citecolor=blue,        
}

\allowdisplaybreaks

\begin{document}

\title{Multipole order in two-dimensional altermagnets}

\author{Tenta Tani}
\email{tenta.tani@qp.phys.sci.osaka-u.ac.jp}
\affiliation{Department of Physics, The University of Osaka,
Toyonaka, Osaka 560-0043, Japan}

\author{Ulrich Z\"ulicke}
\email{uli.zuelicke@vuw.ac.nz}
\affiliation{MacDiarmid Institute, School of Chemical and
Physical Sciences, Victoria University of Wellington, PO
Box 600, Wellington 6140, New Zealand}

\date{\today}

\begin{abstract}
We theoretically investigate the magnetic-multipole
orders in two-dimensional (2D) altermagnets, focusing on
two representative  models: a generic minimal three-site
model, and a four-site model representative of monolayer
FeSe. We construct low-energy effective Hamiltonians for
both systems and calculate their respective multipole
indicators to characterize the underlying magnetic
order. Our analysis reveals an intriguing contrast
between the two systems. We find that the generic
minimal model exhibits the expected non-zero
magnetic-octupole order. In the monolayer-FeSe model,
however, the magnetic-octupole order vanishes globally,
and a magnetic-hexadecapole order is present instead.
The emergence of altermagnetic splitting in the band
structure then arises via the interplay with a
sublattice-isospin degree of freedom. Our work
demonstrates how the classification and comprehensive
understanding of 2D altermagnetic materials transcends
bulk descriptions.
\end{abstract}

\maketitle

\section{Introduction}

Altermagnetism~\cite{Jungwirth2022,Jungwirth2022-rev,
Roig2024,Radaelli2024,Song2025} has recently been
introduced as a term to designate magnetically ordered
materials that have no bulk magnetization but still
exhibit spin splitting in the electronic band structure.
Altermagnets are thus different from ferromagnets (which
have a finite spontaneous magnetization and associated
uniform exchange spin splitting of electron states) and
the more ordinary antiferromagnets (which also lack a bulk
magnetization but retain spin degeneracy in their band
structure because time inversion combined with a
translation or with space inversion remains a good
symmetry).

Physical ramifications of altermagnetism are currently
attracting considerable attention. The altermagnetic
properties of several materials have been intensively
investigated, including $\rm RuO_2$~\cite{Smejkal2023,
Lovesey2023,Bai2023,Zhou2024,Fedchenko2024},
MnTe~\cite{Mazin2023,Lee2024,Aoyama2024}, and $\rm
CrSb$~\cite{Ding2024,Reimers2024,Zhou2025}. For example,
spin-to-charge conversion~\cite{Bai2023}, thermal
transport~\cite{Zhou2024}, and
magnons~\cite{Smejkal2023} have been examined for $\rm
RuO_2$. Two-dimensional (2D) altermagnets are also
increasingly coming into focus~\cite{Brekke2023,
Sodequist2024,Zeng2024,Asgharpour2025,Mavani2025}, with
proposed materials examples including $\rm
MnPSe_3$~\cite{Mazin2023FeSe}, $\rm
Cr_2SO$~\cite{Guo2023} and $\rm V_2SeTeO$~\cite{Zhu2024}. 

Conceptually, altermagnetism in bulk materials can be
understood as a manifestation of hidden multipole
order~\cite{Hayami2019,Winkler2023,Bhowal2024,
Radaelli2024} that is space-inversion symmetric but
breaks time inversion. In contrast to ferromagnets that
are characterized by a macroscopic magnetic-dipole
density (the magnetization), altermagnets have been shown
to exhibit spontaneous ordering of higher-odd-rank
magnetic multipoles (octupoles, triakontadipoles, etc.)
\cite{Winkler2023,Bhowal2024,Radaelli2024} or
even-rank magnetotoroidal multipoles~\cite{Radaelli2024,
Winkler2024}.

\begin{figure}[b]
\includegraphics [height=3.7cm]{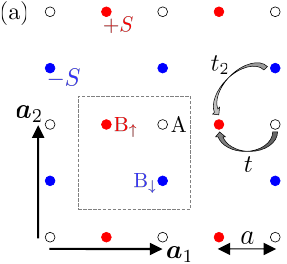}\hfill
\includegraphics [height=3.7cm]{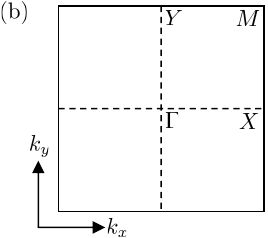}
\caption{\label{fig:squarelattice}%
Minimal model for a 2D altermagnet~\cite{Brekke2023}.
(a)~Square lattice with sites having a localized spin-up
(red), spin-down (blue), or no magnetization (white).
Primitive lattice vectors $\vb*{a}_1, \vb*{a}_2$ are
shown as black arrows, and a (next-)nearest-neighbor
hopping process with parameter $t$ ($t_2$) is depicted
by gray arrows. (b)~First Brillouin zone of the 2D
square lattice.}
\end{figure}

\begin{figure}[t]
\includegraphics [width=1.0\linewidth]{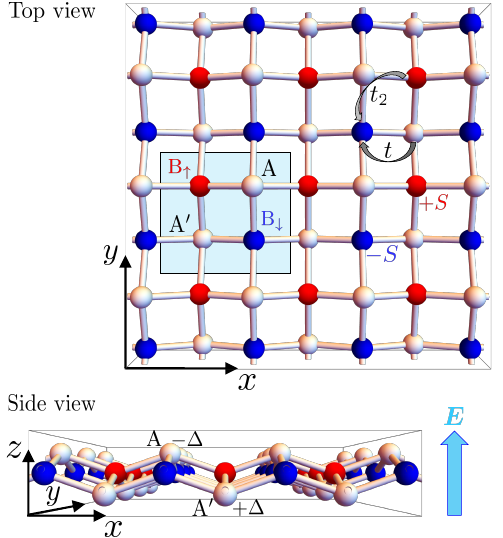}
\caption{\label{fig:FeSeModel}%
Structure representing monolayer FeSe. The unit cell
(indicated by the cyan square) contains two Fe atoms (red
and blue) and two Se atoms (white). In the tight-binding
model, nearest-neighbor ($t$) and second-nearest-neighbor
($t_2$) hopping parameters are included. The Fe sites are
assumed to have localized spins $\pm S$. By applying
a perpendicular electric field $\vb*{E}$, the two Se
sites acquire different potentials $\pm\Delta$.}
\end{figure}

In the present work, we investigate the connection to
multipole order for low-dimensional altermagnet
realizations. For this purpose, we consider two specific
microscopic models. The first one is the minimal model for
a 2D altermagnet introduced in Ref.~\cite{Brekke2023} and
illustrated in Fig.~\ref{fig:squarelattice}(a), with two
magnetic atoms and an additional nonmagnetic atom per
unit cell. The second model is representative of
monolayer FeSe, which was shown to become an altermagnet
in the presence of a perpendicular electric
field~\cite{Mazin2023FeSe}. See
Fig.~\ref{fig:FeSeModel} for the corresponding crystal
structure, which can be regarded just as the minimal
model~\cite{Brekke2023} displayed in
Fig.~\ref{fig:squarelattice}(a) with one additional
nonmagnetic atom in the unit cell.

Our theoretical descriptions build upon previous
foundational work: the minimal model for 2D
altermagnetism proposed in Ref.~\cite{Brekke2023} and the
density-functional-theory (DFT) study of monolayer FeSe
\cite{Mazin2023FeSe}. However, these prior studies did not
investigate the nature of the magnetic-multipole orders
that are intrinsically linked to the altermagnetic phases
in these systems. The primary goal of this work is to
fill this gap. To this end, we make two key
contributions: (i) For the model from
Ref.~\cite{Brekke2023}, we derive the effective
low-energy Hamiltonian and identify the characteristic
magnetic-octupole order. (ii) For monolayer FeSe, we
construct a tight-binding model that captures the
essential low-energy features of the DFT
results~\cite{Mazin2023FeSe}, and we demonstrate that this
system hosts a hidden magnetic-hexadecapole order that is
linked to the altermagnetic spin splitting. Through these
analyses, we clarify the essential role of multipole
physics in these prototypical 2D altermagnets.

We treat each model by first diagonalizing their
respective tight-binding Hamiltonians that include
nearest-neighbor and next-nearest-neighbor hopping terms,
as well as local exchange-energy splittings at the
magnetic sites. (For the case of monolayer FeSe, the
electrostatic potential arising from the perpendicular
electric field is also included.) This yields the known
altermagnetic band structures for the two model
systems~\cite{Brekke2023,Mazin2023FeSe}. We then derive a
faithful effective envelope-function-Hamiltonian
description for the low-energy subbands by adapting the
standard subband-$\vb{k}\cdot\vb{p}$ theory that was
originally devised for treating 2D-electron states in
semiconductor heterostructures~\cite{Broido1985,
Yang1985}. We identify terms in the effective low-energy
model Hamiltonian that are representative of the hidden
multipole order and derive the associated band-structure
indicators using the method proposed in Sec.~II~E of
Ref.~\cite{Winkler2023}. This enables quantification of
higher-order magnetic-multipole densities in the 2D
altermagnetic structures considered here. Our analytical
results reveal intriguing parametric dependencies of the
multipole order.

The remainder of this Article is organized as follows.
Section~\ref{sec:altermag} is devoted to analyzing the
minimal model for 2D altermagnets proposed in
Ref.~\cite{Brekke2023} and shown in
Fig.~\ref{fig:squarelattice}(a). We derive the effective
low-energy effective model for the band structure and
obtain the indicator of the magnetic-octupole density
that is associated with altermagnetism in this structure.
The magnitude of the latter is then calculated both
analytically and numerically. In Sec.~\ref{sec:fese}, we
consider the simple model for 2D FeSe depicted in
Fig.~\ref{fig:FeSeModel}, and derive its effective model.
Surprisingly, we find that the magnetic-octupole density
vanishes globally, while a magnetic-hexadecapole density
represents hidden order in the FeSe system. Altermagnetic
band splitting is shown to arise from the intricate
interplay between the hexadecapolar order and the
perpendicular electric field. Finally,
Sec.~\ref{sec:conclusion} presents our conclusions.

\section{Magnetic-octupole order in a minimal model for
2D altermagnets}\label{sec:altermag}

To obtain generic insights into 2D altermagnets, we
consider the minimal model proposed in
Ref.~\cite{Brekke2023}. As shown in
Fig.~\ref{fig:squarelattice}(a), the model consists of a
nonmagnetic site $\rm{A}$ and two magnetic sites
$\rm{B}_\uparrow$ and $\rm{B}_\downarrow$. The unit cell
is indicated by the dashed square, and the primitive
lattice vectors are $\vb*{a}_1 = (2 a, 0)$, $\vb*{a}_2 =
(0, 2 a)$. In momentum space, reciprocal-lattice vectors
are given by $\vb*{b}_1 = (\pi/a, 0)$, $\vb*{b}_2 =
(0, \pi/a)$. In Fig.~\ref{fig:squarelattice}(b), the
first Brillouin zone with labelled high-symmetry points
is shown.

\begin{figure}[t]
\includegraphics [width=0.95\linewidth]{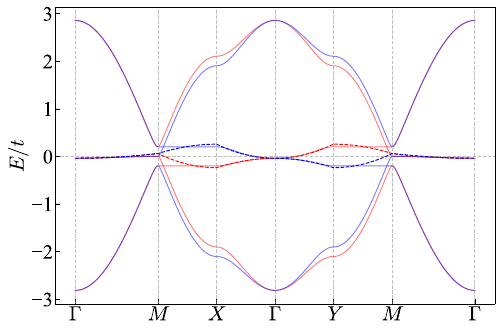}
\caption{\label{fig:band}
Band structure for the minimal 2D-altermagnet model.
The six bands plotted as the solid lines are calculated
from the original Hamiltonian \eqref{eq:original-hamil}.
The dispersions shown as dashed lines are obtained from
the effective Hamiltonian \eqref{eq:eff-hamil}. The red
(blue) color indicates spin-up (spin-down) bands. The
hopping parameters are set to be $J/t=0.2,t_2/t=0.01$.}
\end{figure}

The electron Hamiltonian of the model is given by
\begin{equation}
\begin{split}
\mathcal{H} = \sum_{\sigma=\pm}
\Big( t \sum_{\langle i,j \rangle} c^\dagger_{i,\sigma}
c_{j,\sigma} &+ t_2 \sum_{\langle\langle i,j \rangle
\rangle} c^\dagger_{i,\sigma} c_{j,\sigma} \\[0.1cm]
&\hspace{0.3cm} - J_{\rm sd}\, \sigma \sum_i S_i\,
c^\dagger_{i,\sigma} c_{i,\sigma}\Big) \,\, ,
\end{split}
\end{equation}
where $c^\dagger_{i,\sigma}$ and $c_{i,\sigma}$ are the
creation and annihilation operators of an electron with
spin $\sigma=\pm 1$ at lattice site $i$. The parameter
$t$ quantifies hopping between nearest-neighbor pairs
$\langle i,j \rangle$, and $t_2$ is the corresponding
hopping parameter between second-nearest-neighbor sites
$\langle\langle i,j \rangle\rangle$. Examples for these
hopping processes are illustrated in
Fig.~\ref{fig:squarelattice}(a). The prefactor $J_{\rm
sd}$ of the third term is the exchange coupling between
localized and itinerant electrons. The respective
localized-electron spins of atoms on the $\rm{A}$,
$\rm{B}_\uparrow$, and $\rm{B}_\downarrow$ sites are
given by $S_i = 0$, $S$, and $-S$, respectively. Taking
Bloch states $\{ \ket{A}, \ket{B_\uparrow},
\ket{B_\downarrow} \}$ as the basis, the Bloch
Hamiltonian is written as
\begin{equation}\label{eq:original-hamil}
H(\vb*{k},\sigma) = \mqty(0 & 2\cos{k_x} & 2\cos{k_y} \\
2\cos{k_x} & -J\sigma & 4t_2\cos{k_x}\cos{k_y} \\
2\cos{k_y} & 4t_2\cos{k_x}\cos{k_y} & J\sigma) ,
\end{equation}
where $J=J_{\rm{sd}}S$, and we set $a=1$, $t=1$ for
simplicity. From this Hamiltonian, the electronic band
structure is obtained as shown in Fig.~\ref{fig:band} by
the solid curves, using parameters $J/t = 0.2$ and $t_2/t
= 0.01$. Red (blue) curves are for the spin-up (down)
sector; these exhibit the altermagnetic spin splitting.

\begin{figure}[t]
\begin{center}
\includegraphics [width=0.85\linewidth]{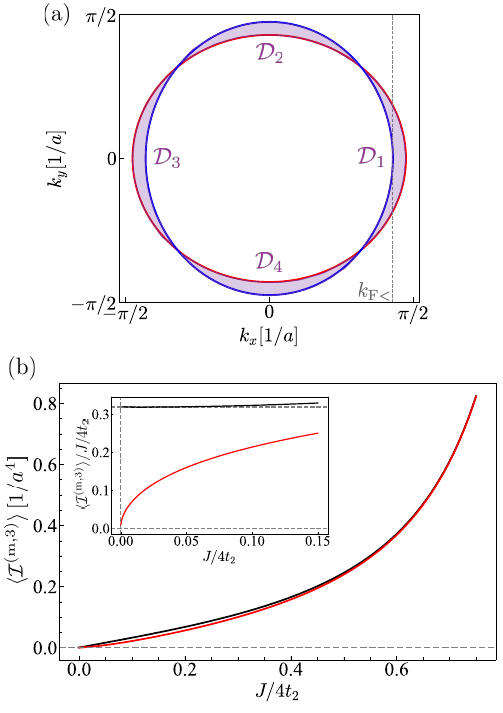}
\caption{\label{fig:expec-value}
(a)~Fermi surfaces for spin-up (red) and spin-down
(blue) states for the effective low-energy Hamiltonian
\eqref{eq:eff-hamil} derived for the generic
2D-altermagnet model~\cite{Brekke2023} with parameters
$J / 4t_2 = 0.1$. Purple shading indicates the four
regions with spin-polarized occupation. (b)~Expectation
value of the magnetic-octupole indicator $\langle
\mathcal{I}^{(\rm{m},3)}\rangle$. The black (red) curve
plots the numerical (analytical) result.}
\end{center}
\end{figure}

We now focus on the low-energy spectrum around the
$\Gamma$ point. An effective two-band Hamiltonian
$H_{\rm{eff}}$ describing states near the Fermi level is
constructed as follows. Exactly at the $\Gamma$ point,
the Hamiltonian~\eqref{eq:original-hamil} becomes
\begin{equation}
H_{\Gamma}(\sigma) = \mqty(0 & 2 & 2 \\
2 & -J\sigma & 4t_2 \\
2 & 4t_2 & J\sigma) \,\, .
\end{equation}
If we set $t_2=0$, $H_\Gamma$ can be analytically
diagonalized. Assuming further $J\ll1$ and neglecting
$\mathcal{O}(J^2)$ terms, the eigenenergies are $E_1=0$,
$E_2 = -2\sqrt{2}$ and $E_3 = 2\sqrt{2}$. We transform
the original Hamiltonian~\eqref{eq:original-hamil} to
this band basis. The unitary matrix $U$ for this
transformation reads
\begin{equation}
U = \frac{1}{4}
\mqty(-\sqrt{2}J\sigma & -2\sqrt{2}-J\sigma & 2\sqrt{2}
-J\sigma \\
-2\sqrt{2} & 2+\sqrt{2}J\sigma & 2-\sqrt{2}J\sigma \\
2\sqrt{2} & 2 & 2) \,\, .
\end{equation}
Hence, the Hamiltonian in the new basis is obtained by
$H'(\vb*{k},\sigma) = U^\dagger H(\vb*{k},\sigma) U$,
which can be written as
\begin{widetext}
\begin{equation}
\begin{split}
&H'(\vb*{k},\sigma) =\\
&\mqty(-4t_2C_xC_y + J\sigma(C_x-C_y) & C_x-C_y+
\frac{J\sigma}{\sqrt{2}}(1-C_y) & -C_x+C_y+
\frac{J\sigma}{\sqrt{2}}(1-C_y)\\
C_x-C_y+\frac{J\sigma}{\sqrt{2}}(1-C_y) & -\sqrt{2}(C_x+
C_y)+2t_2C_xC_y-\frac{J}{2}(3C_x+C_y) & 2t_2C_xC_y +
\frac{J}{2}(C_x - C_y)\\
-C_x+C_y+\frac{J\sigma}{\sqrt{2}}(1-C_y) & 2t_2C_xC_y +
\frac{J}{2}(C_x - C_y) & \sqrt{2}(C_x+C_y)+2t_2C_xC_y-
\frac{J}{2}(3C_x+C_y) )\, ,
\end{split}
\end{equation}
where we used the abbreviations $C_x \equiv \cos{k_x}$
and $C_y \equiv \cos{k_y}$. When we further neglect
$\mathcal{O}(J)$ and $\mathcal{O}(t_2)$ corrections to
terms of $\mathcal{O}(1)$, we find
\begin{equation}\label{eq:transformed-hamil-divided}
H'(\vb*{k},\sigma) = 
\mqty(-4t_2C_xC_y & C_x-C_y & -C_x+C_y \\
C_x-C_y & -\sqrt{2}(C_x+C_y) & 2t_2C_xC_y \\
-C_x+C_y & 2t_2C_xC_y & \sqrt{2}(C_x+C_y))+J\sigma
(C_x-C_y)\mqty(1&0&0 \\ 0&0&1/2 \\ 0&1/2&0).
\end{equation}
\end{widetext}
In Eq.~\eqref{eq:transformed-hamil-divided}, the coupling
between the $E=\mathcal{O}(t_2)\approx 0$ band and the
bands with energies $|E|\approx 2\sqrt{2}$ is negligible
in the vicinity of the $\Gamma$ point, and the effective
low-energy Hamiltonian is found as
\begin{equation}\label{eq:eff-hamil}
\begin{split}
H_{\rm{eff}} (\vb*{k}) &= J\sigma_z(\cos{k_x}-\cos{k_y})
- 4t_2\cos{k_x}\cos{k_y} \\
&\approx -\frac{J}{2}\,\sigma_z(k_x^2-k_y^2) + 4t_2
(\vb*{k}^2/2 - 1) \,\, .
\end{split}
\end{equation}
Here $\sigma_z$ denotes the Pauli matrix in spin space.
The band structure obtained from this effective
Hamiltonian is shown by the dashed curves in
Fig.~\ref{fig:band}, where the red (blue) color
corresponds to the spin-up (spin-down) electron states.
We find that the effective bands fit the original
dispersions fairly well and reproduce the altermagnetic
spin-splitting.  The spin-dependent Fermi surface of the
model is depicted in Fig.~\ref{fig:expec-value}(a), for
$J<4t_2$, which has been called the weak-altermagnet
phase~\cite{Das2024,fu2025}. The shapes of the Fermi
surfaces are ellipses with a semi-major axis $k_{\rm{F}>}
= [2/(1+J/4t_2)]^{1/2}$ and a semi-minor axis $k_{\rm{F}<}
= [2/(1-J/4t_2)]^{1/2}$.

The connection between altermagnetism and multipole
order is embodied in the spin-splitting term $\propto
\sigma_z(k_x^2-k_y^2)$ in the
Hamiltonian~\eqref{eq:eff-hamil}. Specifically, the
quantity
\begin{equation}\label{eq:octupoleI}
\mathcal{I}^{(\rm{m},3)}(\vb*{k}, \sigma) = \sigma
(k_x^2-k_y^2)
\end{equation}
constitutes the band-structure indicator for a
magnetic-octupole density~\cite{Winkler2023,Winkler2024}
that is compatible with the system's magnetic-point-group
symmetry $D_{4h}(D_{2h})$~\cite{Brekke2023}. The
indicator's expectation value is straightforwardly
calculated as
\begin{equation}\label{eq:expec-value}
\begin{split}
\langle \mathcal{I}^{(\rm{m},3)} \rangle &=
\sum_\sigma \int_{\rm{BZ}} \frac{d\vb*{k}}{(2\pi)^2} 
\,\,\, \mathcal{I}^{(\rm{m},3)}(\vb*{k}, \sigma)\,\,
f(E_{\vb*{k}\sigma}) \\
&= \int_{\rm{BZ}} \frac{d\vb*{k}}{(2\pi)^2}\,\,
(k_x^2 - k_y^2)\, \left[ f(E_{\vb*{k}+}) -
f(E_{\vb*{k}-}) \right] ,
\end{split}
\end{equation}
where $f(E)$ is the Fermi distribution function and
$E_{\vb*{k}\sigma}$ the band dispersions close to the
Fermi energy.

The expectation value $\langle\mathcal{I^{(\rm{m},3)}}
\rangle$ directly quantifies the magnetic-octupole
density present in the system. (This is entirely
analogous to how a finite expectation value of
$\mathcal{I}^{(\rm{m},1)} = \sigma$ quantifies the
magnetization. Obviously, $\langle
\mathcal{I^{(\rm{m},1)}}\rangle\equiv 0 $ in the present
case.) We now calculate $\langle\mathcal{I^{(\rm{m},3)}}
\rangle$ in the zero-temperature limit, where the
difference of Fermi functions in the expression
\eqref{eq:expec-value} is finite only for states 
occupied by either spin-up or spin-down electrons.
Considering the shape of the Fermi surface
[Fig.~\ref{fig:expec-value}(a)], the four purple regions
$\mathcal{D}_1$ to $\mathcal{D}_4$ contribute equally.
Thus, Eq.~\eqref{eq:expec-value} can be rewritten as
\begin{equation}\label{eq:expec-value-D}
\langle \mathcal{I}^{(\rm{m},3)} \rangle =
\frac{4}{(2\pi)^2} \int_{\mathcal{D}_1} d\vb*{k}\,\,
(k_x^2-k_y^2) \,\, .
\end{equation}
To obtain an approximate analytical expression, we only
integrate over the region with $k_x > k_{\rm{F}<}$ [see
the gray vertical line in Fig.~\ref{fig:expec-value}(a)],
which is a good approximation for larger $J$. After some
computations, we find
\begin{widetext}
\begin{equation}\label{eq:expec-value-ana}
\langle\mathcal{I}^{(\rm{m},3)}\rangle =
\frac{1}{(2\pi)^2}\left[\left(
\frac{\pi}{2} - \arcsin{\frac{k_{\rm{F}<}}{k_{\rm{F}>}}}
\right) k_{\rm{F}<} k_{\rm{F}>} (k_{\rm{F}>} +
k_{\rm{F}<}) (k_{\rm{F}>} - k_{\rm{F}<}) -
\frac{k_{\rm{F}<}^{2}}{3} \sqrt{ 1 -
\frac{k_{\rm{F}<}^{2}}{k_{\rm{F}>}^{2}} } \left(
k_{\rm{F}<}^{2} - 3k_{\rm{F}>}^{2} +
\frac{2k_{\rm{F}<}^{4}}{k_{\rm{F}>}^{2}}\right)\right]
\,\, .
\end{equation}
\end{widetext}

In Fig.~\ref{fig:expec-value}(b), we compare the
numerical result \eqref{eq:expec-value} (black curve)
and the analytical result \eqref{eq:expec-value-ana}
(red curve). The agreement between both is good for
larger $J/4t_2$, as expected from the approximation
used in the analytical calculation. Deviations become
apparent in the small-$J$ regime ($J/4t_2 \ll 1$), where
we have
\begin{equation}
k_{\rm{F}>}\approx \frac{\sqrt{2}}{a} \left( 1 +
\frac{J/4t_2}{2} \right)\, ,\,
k_{\rm{F}<} \approx \frac{\sqrt{2}}{a} \left(1 -
\frac{J/4t_2}{2}\right)\, ,
\end{equation}
so that Eq.~\eqref{eq:expec-value-ana} reduces to
\begin{equation}\label{eq:expec-value-smallJ}
\langle\mathcal{I}^{(\rm{m},3)}\rangle \approx \frac{16
\sqrt{2}}{3\pi^2 a^4}\left(\frac{J}{4t_2}\right)^{3/2}
\,\, .
\end{equation}
As seen in the inset of Fig.~\ref{fig:expec-value}(b), the
asymptotic behavior of the numerical result is $\langle
\mathcal{I}^{(\rm{m},3)}\rangle \propto J/(4 t_2)$, and
this is not reproduced by the analytical form
\eqref{eq:expec-value-smallJ}. This discrepancy arises
because the area left out of the integration range in
the analytical calculation becomes significant in the
small-$J$ limit.

\section{Altermagnetic monolayer
F\MakeLowercase{e}S\MakeLowercase{e}}\label{sec:fese}

In this section, we investigate monolayer FeSe. Its
crystal structure is shown in Fig.~\ref{fig:FeSeModel}.
It has co-planar magnetic (Fe) sites and a buckling that
renders the two Se sites in the unit cell inequivalent.
As was done in Ref.~\cite{Mazin2023FeSe}, we assume
checkerboard antiferromagnetic order to be present. The
system is then characterized by the magnetic point group
$C_{4h}(S_4)$, which has the combination of space
inversion and time inversion as a good symmetry. As a
result, the band dispersions are spin-degenerate and
asymmetric around the $\Gamma$ point. Application of a
perpendicular electric field was shown to render the
material an altermagnet~\cite{Mazin2023FeSe}.

\begin{figure}[b]
\begin{center}
\includegraphics [width=\linewidth]{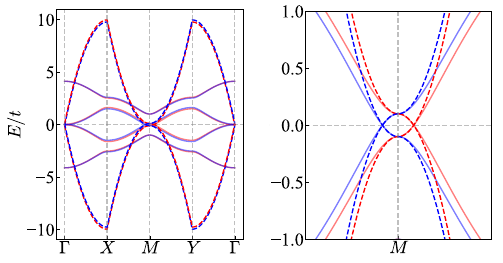}
\caption{\label{fig:band-FeSe}
Band structure of the four-site model for monolayer FeSe.
The eight bands depicted by the solid lines are
calculated from the original Hamiltonian
\eqref{eq:4x4-hamil}, while the four bands shown by the
dashed lines are obtained from the effective Hamiltonian
\eqref{eq:eff-hamil-4x4}. Red (blue) color represents the
spin-up (spin-down) dispersion. The right panel is a
magnification around the $M$ point. Parameters used in the
calculation are $J/t=1$ ,$t_2/t=0$, and $\Delta/t=0.1$.}
 \end{center}
 \end{figure}

We construct a tight-binding model for
checkerboard-antiferromagnetic FeSe simply by adding an
additional nonmagnetic site $\rm{A}'$ to the minimal
2D-altermagnet model discussed in
Sec.~\ref{sec:altermag}. An external perpendicular
electric field is introduced via onsite potentials
$\pm\Delta$ on the Se sites. We thus obtain a $4\times4$
Bloch Hamiltonian,
\begin{equation}\label{eq:4x4-hamil}
\begin{split}
&H(\vb*{k},\sigma) = \\
&\mqty(-\Delta & 0 & 2\cos{k_x} & 2\cos{k_y} \\
0 & \Delta & 2\cos{k_y} & 2\cos{k_x} \\
2\cos{k_x} & 2\cos{k_y} & -J\sigma & 4t_2\cos{k_x}
\cos{k_y} \\
2\cos{k_y} & 2\cos{k_x} & 4t_2\cos{k_x}\cos{k_y} & J\,
\sigma),
\end{split}
\end{equation}
with basis states $\{ \ket{A}, \ket{A'},
\ket{B_\uparrow}, \ket{B_\downarrow} \}$. The band
structure calculated from this Hamiltonian for some
sample parameters is plotted by the solid curves in
Fig.~\ref{fig:band-FeSe}. Around the $M$ point, this
model reproduces the DFT band structure calculated in
Ref.~\cite{Mazin2023FeSe}.

At the $M$ point, the Hamiltonian \eqref{eq:4x4-hamil} is
diagonal with eigenenergies $[E^{(0)}_{1\sigma},
E^{(0)}_{2\sigma},E^{(0)}_{3\sigma},E^{(0)}_{4\sigma}] =
[-\Delta,\Delta,-J\sigma,J\sigma]$. In order to derive a
low-energy effective Hamiltonian, we apply second-order
perturbation theory to these eigenstates. This yields the
energy dispersions
\begin{equation}
E_{\vb*{k}i\sigma} = E_{i\sigma}^{(0)} + \sum_{j\neq i}
\frac{\mel*{u_i^{(0)}}{H'}{u_j^{(0)}}}{E^{(0)}_{i\sigma}-
E^{(0)}_{j\sigma}} \,\, ,
\label{eq:secondorder-perturbation}
\end{equation}
for $i=1,2,3,4$.
$H'$ is defined via $H=H_0+H'$ and $H_0=H(\vb*{k}=M)$.
$\ket*{u_{i}^{(0)}}$ is the $i$th eigenvector of $H_0$.

Here we are interested in the two bands with $i=1,2$,
which represent the band structure closest to the Fermi
energy $E=0$. When we relabel $i=1,2$ in terms of a
pseudospin quantum number $\tau=\pm 1$, we have
\begin{equation}
E_{\vb*{k}\tau\sigma} =
\tau \left[\Delta + 4\frac{(\Delta+J\sigma)\cos^2{k_x}
+ (\Delta-J\sigma)\cos^2{k_y}}{\Delta^2-J^2}\right]\, .
\end{equation}
We can derive an effective Hamiltonian considering the
limit of a small perpendicular electric field. For $J \gg
\Delta$,
\begin{equation}
H_{\rm{eff}}(\vb*{k}) = \tau_z [\Delta - (4/J)\sigma_z
(\cos^2{k_x} - \cos^2{k_y})] \,\, ,
\end{equation}
where $\tau_z$ is the Pauli matrix in the low-energy
two-band space. When we take the $M$ point to be the
origin of $\vb*{k}$ space, we have
\begin{equation}\label{eq:eff-hamil-4x4}
H_{\rm{eff}}(\vb*{k}) = \tau_z  [\Delta - (4/J)\sigma_z
(k_x^2 - k_y^2)] \,\, , 
\end{equation}
for small $\vb*{k}$. Plotting the band dispersions for the
effective Hamiltonian \eqref{eq:eff-hamil-4x4} by dashed
lines in Fig.~\ref{fig:band-FeSe}, we see that they
coincide with the original low-energy bands close to the
$M$ point.

\begin{figure}[b]
\begin{center}
\includegraphics [width=0.85\linewidth]{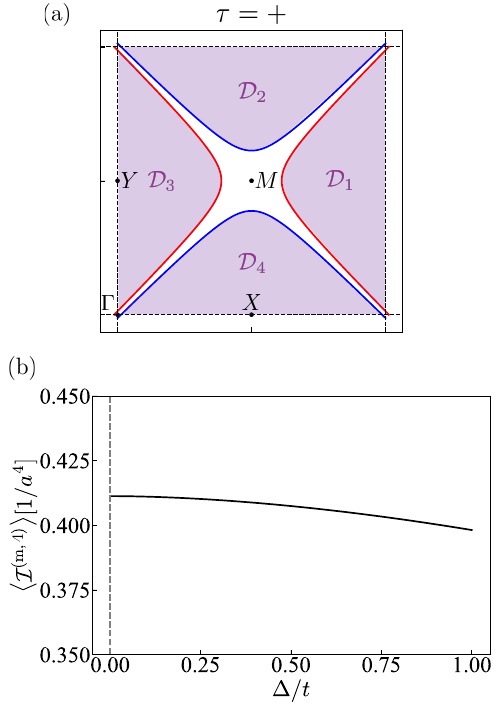}
\caption{\label{fig:expec-value-FeSe}%
(a)~Fermi surfaces of the minimal model for FeSe at the
$M$ point. Here we plot the $\tau=+$ sector only. Regions
$\mathcal{D}_1$ and $\mathcal{D}_3$ ($\mathcal{D}_2$ and
$\mathcal{D}_4$) are occupied by spin-up(-down) states.
Parameters used are $J/t=1$, $t_2/t=0$, and $\Delta/t=
0.5$. (b)~Expectation value $\langle
\mathcal{I}^{(\rm{m},4)} \rangle$ of the
magnetic-hexadecapole indicator \eqref{eq:hexaI},
plotted as a function of the electric-potential
magnitude $\Delta$.}
 \end{center}
 \end{figure}

The Fermi surface calculated from the effective
Hamiltonian \eqref{eq:eff-hamil-4x4} for the $\tau=+$
band is shown in Fig.~\ref{fig:expec-value-FeSe}(a). It
consists of hyperbolas $k_x^2/k_{\rm{F}0}^2 - k_y^2/
k_{\rm{F}0}^2 = \pm1$, where the $+/-$ signs are for the
spin-up/-down electrons, respectively. Here we introduced
the scale $k_{\rm{F}0} = \sqrt{J\Delta}/2$. Spin-up
(spin-down) states are occupied in the region
$\mathcal{D}_1$ and $\mathcal{D}_3$ ($\mathcal{D}_2$ and
$\mathcal{D}_4$). Such a Fermi-surface shape is
indicative of a strong altermagnet~\cite{Das2024,fu2025}.

In the effective Hamiltonian~\eqref{eq:eff-hamil-4x4},
$\sigma_z$ corresponds to the electron spin, whereas
$\tau_z$ is a sublattice-pseudospin degree of freedom
distinguishing the Se sites $\{\rm{A},\rm{A'}\}$.
Therefore, $\tau_z$ has the characteristics of the
spatial coordinate $z$, including its transformational
properties under space inversion (odd) and time
inversion (even). Consequently, the quantity
\begin{equation}\label{eq:hexaI}
\mathcal{I}^{(\rm{m},4)}(\vb*{k},\tau,\sigma) = \tau\,
\sigma\, (k_x^2-k_y^2)
\end{equation}
constitutes a band-structure indicator for a rank-4
multipole that is odd under both space inversion and
time inversion, i.e., a magnetic hexadecapole.

\begin{figure}[t]
\begin{center}
\includegraphics [width=0.8\linewidth]{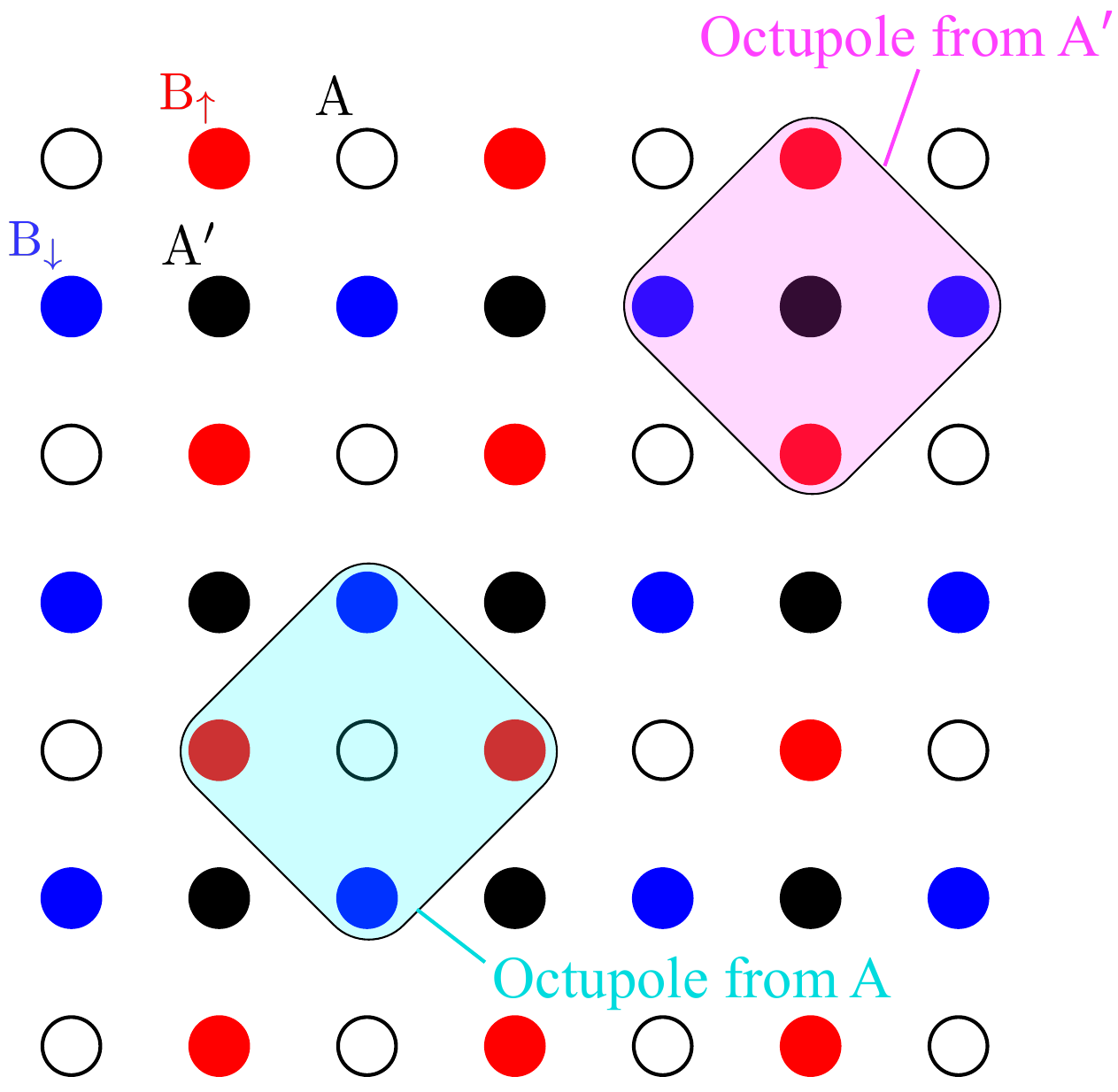}
\caption{\label{fig:octupole-schematic}
Schematic illustration of the atomic structure of
monolayer FeSe with opposite microscopic magnetic-octupole
contributions from the $\mathrm{A}$ and $\mathrm{A}'$
sublattices indicated by the cyan and magenta regions.
The cancellation of the two sublattice's spatially shifted
magnetic-octupole densities produces magnetic-hexadecapole
order.}
\end{center}
\end{figure}

The possibility of real-space quantities, such as
sublattice-related pseudospins, appearing in effective
Hamiltonians and energy dispersions for 2D systems
enlarges the palette of band-structure indicators for
multipole order beyond those that have previously been
constructed for 3D
crystals~\cite{Winkler2023,Winkler2024}.
Magnetic-hexadecapole order is a case in point. In a bulk
crystalline solid, its indicator must be an odd-order
polynomial in wave-vector components~\cite{Winkler2024}
and cannot involve spin. The fact that, for a 2D system,
the combination $\tau_z\,\sigma_z$ transforms like a
wave-vector component enables a substitution $k_j \to
\tau\, \sigma$, leading to the form as given in
\eqref{eq:hexaI}. 
For more explicit illustration of this point, we now
characterize both the magnetic octupole and the magnetic
hexadecapole order in the FeSe model system.

\subsection{Magnetic octupole}
\label{subsec:fese-octupole}

The indicator for a macroscopic magnetic-octupole density
is given in Eq.~\eqref{eq:octupoleI}. Its expectation
value for the FeSe system is obtained by evaluating
\begin{equation}\label{eq:expec-value-4x4}
\langle \mathcal{I}^{(\rm{m},3)} \rangle =
\sum_{\tau\sigma} \, \int_{\rm{BZ}}
\frac{d\vb*{k}}{(2\pi)^2}\,\,\, \mathcal{I}^{(\rm{m},3)}
(\vb*{k}, \sigma)\,\, f(E_{\vb*{k}\tau\sigma})\,\, .
\end{equation}
At zero temperature, Eq.~\eqref{eq:eff-hamil-4x4}
implies $\sum_{\tau\sigma} \, \sigma f(E_{\vb*{k}\tau
\sigma})\equiv \sum_\sigma \, \sigma =0$. Hence, the
integrand from Eq.~\eqref{eq:expec-value-4x4} vanishes
identically, and so does $\langle
\mathcal{I}^{(\rm{m},3)} \rangle$. This result means
that the contributions to a magnetic-octupole density
from the two sublattices $\tau=\pm$ cancel out.
Figure~\ref{fig:octupole-schematic} is a conceptual
illustration of the FeSe atomic structure for an intuitive
explanation of the globally vanishing magnetic octupole,
with the two contributions from the $\mathrm{A}$ sublattice
(cyan) and the $\mathrm{A}'$ sublattice (magenta)
cancelling each other. However, as we now demonstrate, a
magnetic hexadecapole density emerges as the remnant of
this cancellation.

\subsection{Magnetic hexadecapole}

The expectation value of the magnetic-hexadecapole
indicator \eqref{eq:hexaI} for the FeSe system is
obtained from
\begin{equation}\label{eq:expec-value-hexadeca}
\langle \mathcal{I}^{(\rm{m},4)} \rangle =
\sum_{\tau\sigma} \, \int_{\rm{BZ}}
\frac{d\vb*{k}}{(2\pi)^2} \,\,\,
\mathcal{I}^{(\rm{m},4)}(\vb*{k}, \tau, \sigma) \,\,
f(E_{\vb*{k}\tau\sigma})\, .
\end{equation}
In this integral, contributions for $\tau=\pm$ do not
cancel, as we find
\begin{equation}
\sum_{\tau\sigma} \, \tau\,\sigma\, f(E_{\vb*{k}\tau
\sigma}) = 2\, \sum_\sigma\, \sigma\,
f(E_{\vb*{k}+\sigma})\,\, .
\end{equation}
Considering the shape of Fermi surfaces for the $\tau=+$
bands [Fig.~\ref{fig:expec-value-FeSe}(a)], the
expectation value~\eqref{eq:expec-value-hexadeca} is
obtained via
\begin{equation}\label{eq:expec-value-hexadeca-D}
\langle \mathcal{I}^{(\rm{m},4)} \rangle =
\frac{8}{(2\pi)^2} \int_{\mathcal{D}_1} d\vb*{k}\,\,
(k_x^2-k_y^2) \,\, ,
\end{equation}
since the four regions $\mathcal{D}_1$ to
$\mathcal{D}_4$ contribute equally to the result.
We can exactly integrate this to get an expression
\begin{widetext}
\begin{equation}\label{eq:expec-value-ana-FeSe}
\langle\mathcal{I}^{(\rm{m},4)}\rangle =
\frac{k_{\rm{F}0}^4}{2\pi^2}\left[\frac{\pi}{6
k_{\rm{F}0} a}\, \sqrt{\frac{\pi^2}{4 k_{\rm{F}0}^2 a^2}
- 1}\, \left(\frac{\pi^2}{k_{\rm{F}0}^2 a^2} + 2\right)
- 2 \ln( \frac{\pi}{2 k_{\rm{F}0} a} +
\sqrt{\frac{\pi^2}{4 k_{\rm{F}0}^2 a^2} - 1}\, ) \right]
\,\, .
\end{equation}
\end{widetext}
The $\Delta$-dependence of $\langle
\mathcal{I}^{(\rm{m},4)}\rangle$ is plotted in
Fig.~\ref{fig:expec-value-FeSe}(b). Note that it stays
finite in the limit $\Delta=0$, i.e., it is a property
of the FeSe crystal itself. However, the associated
altermagnetic spin splitting emerges only for $\Delta\ne
0$ when the combined spin and sublattice-pseudospin
degeneracy is lifted.
The finite magnetic hexadecapole order could be
experimentally observed via the magneto-piezoelectric
effect, as proposed in Ref.~\cite{Watanabe2017}.

Our finding that the altermagnetic phase in the
monolayer-FeSe model is characterized by a magnetic
hexadecapole looks surprising from the standpoint of
conventional wisdom. Even-rank magnetic multipoles
preserve the combination of space inversion and time
inversion and, thus, have previously been associated with
spin-degenerate, $+\mathbf{k}$/$-\mathbf{k}$ asymmetric
band structures~\cite{Jungwirth2022-rev,Winkler2023} that
are not of altermagnetic form. However, our work reveals a
novel mechanism arising in the 2D FeSe system where the
sublattice degree of freedom provides a crucial coupling
$\tau_z \sigma_z$. This coupling enables the
magnetic-hexadecapole order to generate a spin-dependent
term in the effective Hamiltonian, which in turn drives
the spin splitting. This result thus identifies a new
pathway to altermagnetism, expanding the classes of
magnetic multipole orders known to host this phase.

In principle, the same scenario as discussed here
for a specific 2D system could occur more generally in any
situation with a time-inversion-invariant pseudospin
degree of freedom transforming like an odd-rank polar
tensor. This usually requires band-edge degeneracies
and/or band edges located away from the $\Gamma$ point.

\section{Conclusion}\label{sec:conclusion}

We have investigated multipole order in a minimal model
for generic 2D altermagnets, and for a model system
representing monolayer FeSe. Consistent with crystal
symmetry, the former is shown to exhibit
magnetic-octupole order, whereas the latter hosts a
macroscopic magnetic-hexadecapole density. Both orders
are quantified based on their respective band-structure
indicators, given in Eqs.~\eqref{eq:octupoleI} and
\eqref{eq:hexaI}, respectively.

In contrast to previously considered situations for
3D crystals where magnetic-hexadecapole order is indicated
by band dispersions that are asymmetric in wave-vector
components but always remain spin-degenerate, such order
in a 2D system can give rise to altermagnetic spin
splitting. We link such fundamental differences in
physical ramifications for the same type of multipole
order to the possibility to have 2D band-structure
invariants that involve pseudospins that are equivalent to
real-space (e.g., sublattice) positions. Future work could
focus on incorporating more systematically any possible
pseudospin degrees of freedom into the construction of
band-structure indicators in 2D crystals. 

\begin{acknowledgments}
T.T.\ was supported by JSPS KAKENHI Grant No.\
JP23KJ1497. 
\end{acknowledgments}

\bibliography{Refs}

\end{document}